# Acetonitrile Drastically Boosts Conductivity of Ionic Liquids

Vitaly V. Chaban[Σ1], Iuliia V. Voroshylova[2], Oleg N. Kalugin[3], and Oleg V. Prezhdo[1]

[1] Department of Chemistry, University of Rochester, Rochester, New York 14627, United States

[2] REQUIMTE, Departamento de Química e Bioquímica, Faculdade de Ciências, Universidade do Porto, 4169-007 Porto, Portugal

[3] Department of Chemistry, Kharkiv National University, Kharkiv 61077, Ukraine

We apply a new methodology in the force field generation (*PCCP 2011, 13, 7910*) to study the binary mixtures of five imidazolium-based room-temperature ionic liquids (RTILs) with acetonitrile (ACN). The investigated RTILs are composed of tetrafluoroborate ($BF_4$) anion and dialkylimidazolium cations, where one of the alkyl groups is methyl for all RTILs, and the other group is different for each RTILs, being ethyl (EMIM), butyl (BMIM), hexyl (HMIM), octyl (OMIM), and decyl (DMIM). Specific densities, radial distribution functions, ionic cluster distributions, heats of vaporization, diffusion constants, shear viscosities, ionic conductivities, and their correlations are discussed. Upon addition of ACN, the ionic conductivity of RTILs is found to increase by more than 50 times, that significantly exceeds an impact of most known solvents. Remarkably, the sharpest conductivity growth is found for the long-tailed imidazolium-based cations. This new fact motivates to revisit an application of these binary systems as advanced electrolytes. The ionic conductivity correlates generally with a composition of ionic clusters, simplifying its predictability. In turn, the addition of ACN exponentially increases diffusion and decreases viscosity of the imidazolium-based RTILs/ACN mixtures. Large amounts of acetonitrile stabilize ion pairs, but ruin greater ionic clusters.

[Σ] Corresponding author, Tel: 1-585-276-5751, e-mail: v.chaban@rochester.edu, vvchaban@gmail.com

Textual abstract

Atomistic simulations suggest a sharp enhancement of conductivity of ionic liquids upon solvation by acetonitrile

Graphical Abstract

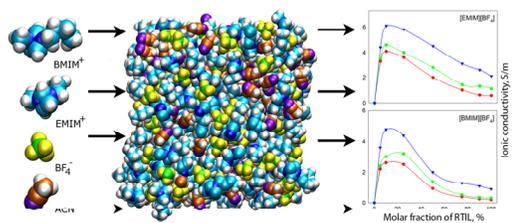

**Introduction**

Room-temperature ionic liquids (RTILs) by the virtue of a complex of their remarkable properties and claimed potential of wide industrial applications have been actively investigated during the past decade[1-4]. These unusual compounds are often positioned as alternatives to the traditional organic solvents, latter exhibiting certain cons including toxicity and flammability. Due to negligible vapor pressure, unique permittivity, high thermal stability, non-flammability, ability to solvate lots of organic and inorganic substances, wide electrochemical window, RTILs are applied as reaction medias[5], separation solvents[6], lubricants[7], novel high-performant and green electrolytes[8], etc. For example, in order to improve safety of conventional lithium batteries, the binary mixtures of lithium salts with ionic liquids have been applied as perspective electrolytes[9].

Stimulated by a steady increasing interest to their application in electrochemistry, various RTILs have been subjected to studies of fundamental electrochemical properties. The electrical conductivity of certain pure RTILs and their mixtures with water has been studied over a wide temperature range[10-13]. However, most pure RTILs are extremely viscous (more than 50 cP at room temperature). As a result, they exhibit quite low self-diffusion ($ca.$ $0.1 \times 10^{-9}$ m$^2$/s) and conductivity ($ca.$ 1 S/m)[14], the latter being a critical value for the performant electrolyte. In order to enhance ionic motion in ionic liquids, water and certain organic solvents can be added as admixtures. The intermolecular and ion-molecular interactions, both mostly electrostatic in nature, are expected to drive dissociation of ion pairs, leading to a greater ionic mobility. A number of experimental and simulation studies[15-17] of the water and RTIL mixtures have been already carried out. In particular, a recent work of Canongia Lopes and coworkers presents a noteworthy cluster analysis, revealing the existence of four distinct structural regimes - isolated water molecules, chain-like water aggregates, bicontinuous system, and isolated ions or small ion clusters[18]. Importantly, the molar conductivity of such systems is found to strongly (exponentially) depend on the molar fraction of the admixture. Similarly, the viscosity of RTILs is reduced by the addition of water. On the other hand, the amount of water in the RTIL/water mixture can also affect reaction rates, selectivity, media polarity, and solvation properties[19-21]. Both cationic and anionic diffusion appreciably grows in the RTIL/water mixtures, thanks to the screening of the electrostatic interactions between water molecules and ions of the ionic liquid.

In contrast to the RTIL/water mixtures, other potential co-solvents for ionic liquids are studied less extensively[17,22-26], although their importance for industrial applications of RTILs is undoubted. Interestingly, the number of experimental studies of this kind of systems significantly exceeds the number of simulation researches[27], drastically differing from the situation for pure RTILs. This observation once again underlines a complexity and abundance of inter-atomic interactions[2,28] in the RTIL containing mixtures.

In the present work, a novel methodology[29,30] to simulate transport properties of RTILs is applied to investigate five RTIL/ACN mixtures. The cations are based on imidazole and contain ethyl, butyl, hexyl, octyl, and decyl groups, whereas the anion (tetrafluoroborate, $BF_4$) is the same for all compounds. Two common imidazolium-based ionic liquids, 1-ethyl-3-methylimidazolium tetrafluoroborate ([EMIM][$BF_4$]) and 1-butyl-3-methylimidazolium tetrafluoroborate ([BMIM][$BF_4$]) with acetonitrile (ACN) (Figure 1) are investigated over the entire range of compositions at 283-323 K. Additionally, three RTILs with longer alkyl tails, namely hexyl ([HMIM][$BF_4$], octyl ([OMIM][$BF_4$]), and decyl ([DMIM][$BF_4$]) (Table 1), which are usually missed in the conductivity-related researches, are employed to derive trends for these binary mixtures. Acetonitrile is a common aprotic solvent with a relatively high diffusion coefficient ($4.3 \times 10^{-9}$ m$^2$/s) and low shear viscosity (0.34 cP). The structure and dynamics of pure ACN are driven by dipole-dipole interactions that predict its fine miscibility with, obviously, polar ionic liquids. Based on the above considerations, we expect that mixing of [EMIM][$BF_4$], [BMIM][$BF_4$], [HMIM][$BF_4$], [OMIM][$BF_4$], and [DMIM][$BF_4$] with ACN should lead to drastic changes in the hindered ionic mobility of these RTILs, resulting in an appropriately increased conductivity. The molecular dynamics (MD) method with pairwise interaction potentials is applied to derive extra-long MD trajectories (0.15-0.25 ms) of the simulated systems. Our recent technique[29] exploiting uniformly scaled electrostatic charges is used to phenomenologically account for an appreciable degree of electronic polarization of RTILs. The available experimental densities and viscosities of the [EMIM][BF$_4$]/ACN and [BMIM][BF$_4$]/ACN at 298 K[14,23,24,31,32] are applied to validate the resulting models. Based on the simulation results, it is demonstrated that the addition the ACN allows for conductivity increase as much as by 1.5 orders of magnitude. We argue that ionic mobility increase occurs thanks to very suitable interactions between ions and ACN, therefore the resulting conductivities only insignificantly depend on the cation size, shape, and mass. An ability to enlarge ionic conductivity of RTIL by varying exclusively the molar fraction of the molecular component favors applications of the imidazolium-based RTILs/ACN as novel non-aqueous electrolytes[33-35].

**Methodology**

The phase trajectories for 16 systems, containing 0, 5, 10, 25, 50, 75, 90, 100 molar percents of [EMIM][BF$_4$] and [BMIM][BF$_4$], and 12 systems, containing 5, 10, 25, 100 molar percents of [HMIM][BF$_4$], [OMIM][BF$_4$], and [DMIM][BF$_4$], each simulated at 283, 298, and 323 K, are derived using classical molecular dynamics simulations. Detailed decription of the simulated systems is summarized in Table 1. The molecular dynamics runs are accomplished using the home-made version of the GROMACS 4.0 simulation package[36] in the constant temperature constant pressure (NPT) ensemble. The pairwise interaction potentials are applied to treat all the molecular, ionic, and ion-molecular interactions in the systems. All atoms of 1-ethyl-3-methylimidazolium tetrafluoroborate, 1-butyl-3-methylimidazolium tetrafluoroborate, 1-hexyl-3-methylimidazolium tetrafluoroborate, 1-octyl-3-methylimidazolium tetrafluoroborate, 1-decyl-3-methylimidazolium tetrafluoroborate, and acetonitrile (Figure 1) are represented by separate interaction centers, possessing Lennard-Jones (12,6) parameters and electrostatic charges. The classical Lorentz-Berthelot rules in the form, suggested for the AMBER-based force fields[37-39], are employed to generate the cross-parameters in the Lennard-Jones equation. The constant temperature of 283, 298, and 323 K is maintained using the velocity rescaling thermostat with a response time of 1.0 ps. The constant pressure (1 bar) is maintained using the Parrinello-Rahman technique for pressure coupling with a relaxation time of 4.0 ps. The widely used leap-frog algorithm is used to integrate the equations of motion with a time-step of 1 fs, whereas the list of the nearest neighbors was updated every 10 time-steps within a sphere of radius of 1.5 nm.

An obviously slow dynamics of particles in RTILs requires longer relaxation and production stages than it is usually performed in conventional dynamics studies (100ps-10,000 ps). It is also necessary for the mixtures with small molar fractions of the ionic liquid, since lower amount of particles (ions) requires longer simulation times to achieve ergodicity. In the present study, the initial relaxation is carried out at 350 K during 4,000 ps in the constant volume constant temperature (NVT) ensemble. Next, the generated systems are gradually cooled down to the target temperature (293, 298, and 323 K) during 2,000 ps. The transport properties under discussion are derived from the 50,000 ps trajectories for each system. Three to five consequent MD trajectories of 50,000 ps are used to obtain reliable averages for ionic conductivity and shear viscosity. To ensure extensive trajectory sampling, each productive run begins with an assignment of random velocities to ions and molecules, *i.e.* the generated trajectories are statistically independent. Therefore, they can be treated using the conventional statistical procedures for such values.

The simulated systems of the mixtures and pure components (Table 1) are placed into cubic MD boxes (Figure 1) with periodic boundary conditions applied along all three Cartesian directions to represent a bulk liquid. The long-range electrostatic forces are treated by the Particle Mesh Ewald method as implemented in Ref.[40] with the cut-off distance for real-space component equal to 1.5 nm. The continuity of the Lennard-Jones potential is assured by using the shifted force method with a switch region between 1.2 nm and 1.3 nm.

### Force fields

A number of force fields (FFs) for RTILs, including imidazolium-based ionic liquids, have been suggested[27,41,42] during the last decade. A broad spectrum of species are covered by the solid developments of Lopes and Padua[43] based on the well-established OPLS/AA methodology and electrostatic charges derived from an electrostatic potential around ions. In turn, Borodin suggested a polarizable FF using the Thole-type approach[42], and these efficient models have already obtained a wide application in the recent computational studies. Recently, it was demonstrated[29,44,45] that the realistic ionic transport of imidazolium-based RTILs can be simulated by means of a uniform decrease of the Coulombic interaction energy between all interaction sites of both ions. It should be noted that the neutral part of the alkyl tail, in general, does not require scaling in order to preserve the compatibility with the force fields for alkanes. However, the corresponding electrostatic charges are so insignificant, that scaling does not modify any observed transport properties[44]. In the case of RTILs with larger tails (6, 8, 10 carbon atoms), the charges of the atoms, comprizing neutral part, should not be altered, providing a compatibility with the existing force fields for alkanes. An ability to account for polarizability without introducing polarizable models greatly saves computational resources needed to achieve an extensive sampling of the collective transport properties (viscosity, conductivity, *etc*). It is, therefore, adopted in the present work.

The last published six-site model of Nikitin[46] is used to simulate acetonitrile. The model correctly reproduces density (773 kg/m$^3$), heat of vaporization (33.5 kJ/mol) structural distributions, and shear viscosity (~0.4 cP) of bulk ACN, as well as the corresponding properties of the acetonitrile/water mixtures. However, diffusion constant at 298 K is noticeably underestimated ($3.4 \times 10^{-9}$ m$^2$/s) as compared to the experimental value ($4.3 \times 10^{-9}$ m$^2$/s). This observation should be taken into account, when the transport properties of the RTIL/ACN mixtures are described quantatively.

Since both the model of RTIL and the model of ACN utilize Lennard-Jones parameters borrowed from the AMBER force field, certain combination rules for Lennard-Jones interactions and scaling 1-4 interactions are prescribed by the original FF. For generality and consistence, the bonded interaction parameters are also transferred from AMBER[37,38].

**Calculated properties**

We derive specific density, radial distribution functions (RDFs), heat of vaporization, cluster size distributions, and transport properties (diffusion constant, shear viscosity and ionic conductivity), writing down the atomic coordinates and interaction energies every 0.02 ps (20 time-steps) during 50,000 ps of each successive simulation. Whereas ionic conductivity, shear viscosity, and cluster size distributions require an entire trajectory for reliable estimation, diffusion constants are estimated using the last 3,000 ps of each of the consequent MD runs. In turn, all other properties can be derived using significantly poorer sampling (1,000 ps) of the equilibrated systems.

Density of the RTILs/ACN mixtures (Figures 2a, 3) and its fluctuation during the simulation are estimated from the oscillations of the MD box volume in the NPT ensemble. Excess molar volumes (Figure 2b) are derived as a difference between the molar volume of the particular mixture and a sum of the molar volumes of pure components, $V_m^{excess} = V_m(\text{ACN/RTIL}) - x(\text{RTIL}) \cdot V_m(\text{RTIL}) - x(\text{ACN}) \cdot V_m(\text{ACN})$. Shear viscosity, $\eta$, (Figures 4, 9, Tables S1-S2) is obtained by integrating the autocorrelation function of the off-diagonal elements of pressure tensor. This method allows for the viscosity of the system to be found using the equilibrium dynamics method, although extensive sampling is required due to the huge pressure variations for a relatively small MD system. Ionic conductivity, $\sigma$, (Figures 5, 9, Tables S1, S2) is obtained in the framework of the Einstein-Helfand formalism from the linear slope of mean-square displacements of the collective translational dipole moment. Importantly, for this method to perform correctly, periodic boundary conditions should be removed prior to computation, *i.e.* "free" ionic diffusion is necessary. Diffusion constant (Figures 6-8), $D$, is computed *via* the Einstein relation, through plotting mean square displacements of all atoms of each species. The corresponding formulas for the transport properties are summarized in Ref. [29]. Heat of vaporization (Figure 10), $H_{vap}$, is estimated at 298 K only. Similar to the case of conventional liquids, vapor phase of RTILs is assumed to contain only neutral ion pairs, [EMIM][BF$_4$] and [BMIM][BF$_4$], which do not interact with one another. The $H_{vap}$ of the mixtures of varying composition is calculated according to the following formula,

$$H_{vap}(\text{RTIL/ACN})_x = U(\text{RTIL/ACN})_x - x \cdot U(\text{ion pair}) + RT , \qquad (1)$$

where $x$ is a molar fraction, $H_{vap}(\text{RTIL/ACN})_x$ is a heat of vaporization of the RTIL/ACN mixture of a given composition, $U(\text{RTIL/ACN})_x$ is a total interaction energy of the mixture of a given composition, and $U(\text{ion pair})$ is a total interaction energy of the isolated ion pair.

Radial distribution functions (Figure 11), $g_{ij}(r)$, are calculated using a classical definition, exploiting the MD trajectory parts of 1,000 ps. Cluster analysis (Figures 12-13) is based on the single linkage approach. It postulates that two structures form a cluster, once they exhibit at least one direct contact. Next, the third structure belongs to this cluster if it has at least one direct contact with any of these two structures. At each trajectory frame, the iterations persist until no more structures can be added to the cluster. In this study, we assume that two ions form a cluster if the distance between the carbon (CR) site of the cation and the boron (B) site of the anion (see Figure 11, inset for designations) does not exceed 0.500 nm and 0.512 nm (Figure 11) for [EMIM][BF$_4$] and [BMIM][BF$_4$], respectively. The above criteria are selected as positions of the first minima on the corresponding RDFs. Despite there are a few sites of strong cation-anion interactions (*e.g.* imidazole hydrogens-fluorine, each of three imidazole carbons-fluorine, methyl carbon-fluorine), our selection, CR-B, is expected to include these pairs implicitly.

### Results and Discussion

The mixture of one of the considered imidazolium-based RTILs, 1-butyl-3-methylimidazolium tetrafluoroborate, with acetonitrile has been studied[41] a few years ago. Whereas this work provides useful data on the structural, thermodynamics, and volumetric properties, the reported ionic transport (diffusion constants) is not reliable, since it is calculated using a sub-diffusion part of the trajectory (200 ps). As well, shear viscosities and ionic conductivities were not reported. In this work, diffusion constants, shear viscosities, and ionic conductivities are systematically discussed, for the first time, using a comprehensive set of the imidazolium cations.

The dependence of specific density of both 1-ethyl-3-methylimidazolium tetrafluoroborate and 1-butyl-3-methylimidazolium tetrafluoroborate upon the molar fraction of RTIL (Figure 2a) is in a satisfactory agreement with the experimental values reported in Ref. [23]. For instance, for 50% [EMIM][BF$_4$] / 50% ACN and 50% [BMIM][BF$_4$] / 50% ACN mixtures, the simulated densities at 298 K are 1137 and 1094 kg/m$^3$, respectively, whereas the experimental values are 1167 and 1120 kg/m$^3$, respectively. Provided that the fluctuations of density for this size of systems during MD simulation is *ca.* 10 kg/m$^3$, the accuracy of the simulated data is good to excellent. The underestimation of the average density by *ca.* 2% comes from the underestimation of density of pure RTILs. The reasons are discussed in Ref. [29] The mixtures, containing [EMIM][BF$_4$], are slightly denser, because the lyophobic alkyl tail of [BMIM][BF$_4$] participates in the ion-molecular and ion-ionic interactions to a lesser extent than the cationic ring. Another explanation is that [EMIM][BF$_4$] contains the largest fraction of fluorine atoms, which are the heaviest species in the studied systems.

Interestingly, at $x$ (RTIL) < 25%, the difference between densities of [EMIM][BF$_4$] and [BMIM][BF$_4$] vanishes. In this context, it should be underlined that density usually drastically affects the molecular and ionic motion, and therefore, conductivity of the electrolyte. Provided that the size, mass and shape of EMIM$^+$ and BMIM$^+$ are similar, their diffusivities at $x$ (RTIL) < 25% are expected to be very close. They are determined by the mobility of the ACN molecules at given temperature.

The excess molar volumes (Figure 2b) of the mixtures are a convenient measure of non-ideality. Over the entire composition range, the deviations are negative. The observed behavior indicates that mixing brings more favorable attractive interactions between RTILs and ACN than in pure components. Both [EMIM][BF$_4$]/ACN and [BMIM][BF$_4$]/ACN mixtures exhibit a minimum at *ca.* 30%, which excellently coincides with the previous experimental observation and MD simulation.[41] The smallest negative deviations are found at 283 K, whereas the largest ones are at 298 K. Generally speaking, the value of $V_m^{excess}$ determines a miscibility of components at given ambient conditions. The behavior of $V_m^{excess}$ at $x$ ([EMIM][BF$_4$]) > 75% at three temperatures provides a new understanding of the interactions in the RTIL-rich mixtures. In particular, adding small amounts of ACN to RTIL is less favorable than adding small amounts of RTIL to ACN (Figure 2b).

The temperature dependence of density is linear, $\rho = \rho_0 + a \cdot T$, in spite of the particular composition of the mixture and ionic species. The constants $\rho_0$ and $a$ obtained using linear regression are depicted in Figure 3. The correlation coefficient exceeds 0.999 for all sets of the fitted data. The density at zero absolute temperature, $\rho_0$, reflects a hypothetic density of the mixtures, assuming that they exist in a liquid state at $T = 0$ K. It is systematically higher for [EMIM][BF$_4$]/ACN and increases as $x$ (RTIL) increases. In turn, $a$ is a derivative of the liquid density with respect to temperature, *i.e.* is equivalent to isobaric expansion coefficient. ACN . On the contrary, $|a|$ decreases as the molar fraction of the ionic liquids increases, but exhibits a non-monotonic behavior as $x$ (RTIL) equals to 5, 75, and 90%. Even though the absolute values of $a$ are relatively small, the observed extrema are statistically meaningful. Their existence suggests that RTILs and ACN form sub-nanometer patterns, which are specific to certain mixture composition. As a result, thermal expansion coefficients exhibit a complicated dependence on mixture composition. Obviously, if the corresponding liquids interacted weakly, $a$ should change smoothly with $x$.

Shear viscosity, $\eta$, (Figure 4) is of great importance for simulation studies of liquids, since it simultaneously reflects molecular and ionic transport and energy dissipation in the investigated systems. Meanwhile, it is affordable for direct physical chemical experiment and plays a major role in the parameterization of certain empirical force fields. Recently, Borodin reported viscosities for a set of pure RTILs at room and elevated temperatures[42]. The viscosities of [EMIM][BF$_4$]/ACN and [BMIM][BF$_4$]/ACN mixtures strongly depend on temperature (Figure 4). This trend is especially pronounced, if the molar fraction of RTIL exceeds 75%. For instance, at $x$ (RTIL) = 90%, $\eta$ = 43 cP (283 K), 22 cP (298 K), 12 cP (323 K) and 137 cP (283 K), 63 cP (298 K), 17 cP (323 K) for [EMIM][BF$_4$]/ACN and [BMIM][BF$_4$]/ACN mixtures, respectively. At the same compositions, the viscosity of [BMIM][BF$_4$]/ACN is systematically and noticeably higher than of [EMIM][BF$_4$]/ACN. The observed difference may contradict an intuition, since density of [BMIM][BF$_4$]/ACN (Figure 2) is somewhat smaller. Meanwhile, a longer lyophobic tail is not expected to enhance negative non-ideal deviations of the mixture properties. Hence, it should not cause an observed viscosity increase as compared to [EMIM][BF$_4$]/ACN. On the other hand, BMIM$^+$ is 1.25 times heavier than EMIM$^+$, that results in a slower ionic motion. It is furthermore strengthened by an energetically favorable solvation of cations by ACN[41]. At 323 K, acetonitrile approaches its normal boiling point ($\eta_{323K}$(ACN) = 0.2 cP) and drastically decreases the viscosity of the entire system (Figure 4). Our data is also well correlated with recent experiments.[26]

Our observations evidence that small admixtures of the aprotic solvents are not enough in order to approach RTIL-rich electrolytes to the conventional electrolytes, nevertheless certain impact is achieved. Naturally, for electrochemical applications of RTILs, the shear viscosity should be decreased down to 5-10 cP to enhance ionic conductivity, which is generally in inverse proportion to $\eta$. This requirement is fulfilled as $x$ (RTIL) is equal and less than 25% (Figure 4). One can admit that at these $x$ (RTIL) ionic dynamics is driven by the ion-molecular interactions, rather than inter-ionic ones.

In most cases, the reported $\eta$s are in a good to excellent agreement with the experimental data provided by Wang *et al.*[23] over the entire range of compositions for both considered RTILs (Table 1). Not only qualitative trends are reproduced, but also the experimental and the simulation values are quite close. Nearly all of the observed deviations can be attributed to the uncertainties of determination. This success encourages us to apply the same force field model to derive other transport properties, *i.e.* diffusion coefficients of each component and ionic conductivities, for which no experimental data are available yet.

The shear viscosity of the system is in inverse proportion to its average diffusion coefficient, $D$, as suggested by the Einstein-Stokes relationship. This relationship holds rigorously, provided that the shape of the particles can be approximated by a sphere. Although this is not the case for imidazolium-based cations, we nevertheless observe the qualitative trend. Figures 5-6 depict the average $D$s of the [EMIM][BF$_4$]/ACN and [BMIM][BF$_4$]/ACN mixtures, as well as the $D$s of all components separately over the entire composition range at 283, 298, and 323 K. All compositional dependences can be well describes by exponential decay analytical functions with three parameters, $D = D_0 + A \times \exp(-B \times D)$, where $D_0$ is a diffusion coefficient of pure component, and $A$ and $B$ are empirical constants. The correlation coefficient in all cases exceeds 0.995, indicating a high accuracy of the calculated data.

Remarkably, the ratio between $D$s of the cation (EMIM$^+$, BMIM$^+$) and the anion changes drastically as $x$ (RTIL) changes from 100 down to 0. In pure RTILs and RTIL-rich mixtures, the diffusion of the anion, $D_-$, constitutes only 60-70 % of the cationic diffusion, $D_+$. This occurs in spite of the fact that cations are appreciably more branchy and weighty, 111 a.m.u. of EMIM$^+$ and 139 a.m.u. of BMIM$^+$, as compared to just 87 a.m.u. of tetrafluoroborate anion. However, the anion contains more polar bonds (B-F), and this hinders its motion across the liquid. On the other hand, $D$ (BF$_4^-$) grows with temperature faster than $D$ (EMIM$^+$) and $D$ (BMIM$^+$)[29]. For instance, at 400 K the discussed diffusion coefficients are comparable and at higher temperature the ratio is expected to invert. The same tendency is observed as the molar fraction of ACN increases. At $x$ (RTIL) = 50% and more, the cationic $D$ is higher; however, at $x$ (RTIL) = 25%, the $D_+/D_-$ ratio inverts. Importantly, the same phenomenon occurs at all three temperatures and for both imidazolium-based RTILs. As $x$ (RTIL) further decreases down to 5%, $D_-$ constitutes 110-130% of the $D_+$. Such a remarkable tendency can be understood in terms of the microscopic structure of the mixtures of each composition and the peculiarities of the ion-molecular interactions in the electrolyte acetonitrile solutions, as follows. First, while the content of the mobile acetonitrile molecules is negligible, EMIM$^+$ (BMIM$^+$) and BF$_4^-$ create very large ionic clusters (Figures 10-11), whose structure is completely assigned by strong long-range Coulombic forces between the counterions. Here, the principal role obviously belongs to the fluorine atoms of the anion and the hydrogen atoms of the imidazole ring. Provided that such microscopic order is established, all anions are tightly bound to the neighboring cations, resulting in the extremely low transport constants (Figures 5-7) and excessive viscosity (Figure 4). As the content of the aprotic component increases, certain amounts of BF$_4^-$ are substituted by the neutral molecules. Although the ACN molecule possesses a high dipole moment (~3.9 D), the bulk self-diffusion of this liquid at the considered temperatures (*ca*. 4-5 $\times 10^{-9}$ m$^2$/s) is at least two orders of magnitude higher than that of pure RTILs (0.01-0.05 $\times 10^{-9}$ m$^2$/s). For instance, in the equimolar mixtures of [BMIM][BF$_4$] and acetonitrile, the average $D$s are 7.5, 13, 24 ($\times 10^{-11}$ m$^2$/s) at 283, 298, and 323 K, that is one order of magnitude higher than the diffusion in pure [BMIM][BF$_4$]. Nearly the same ratio is observed in the case of [BMIM][BF$_4$]/ACN mixtures, where the average $D$s are 14, 22, and 37 ($\times 10^{-11}$ m$^2$/s) at 283, 298, and 323 K. At $x$ (RTIL) < 25%, the large ionic clusters (Figures 10-11) are expected to vanish resulting in an abrupt exponential growth of the diffusion of all components.

Besides, the above observations (Figures 5-6) are important from the practical point of view, since diffusion coefficients oversee the content of RTIL, where the ionic motion is controlled by the aprotic solvent, rather than by RTIL. Figures 5-6 demonstrate that the diffusion of the imidazolium-based RTILs is mostly driven by the inter-ionic interactions, and therefore, can be tuned *via* the modification of composition. On the other hand, the disability of acetonitrile to create strong non-covalent bonds with the ions favors its application as an accelerator of the ionic transport.

As we illustrated above using shear viscosity and diffusion, a drastic increase of the molecular and ionic motion is expected as the molar fraction of RTILs is decreased down to 25% and less. The ionic conductivities, σ, (Figure 7) of both [EMIM][BF$_4$]/ACN and [BMIM][BF$_4$]/ACN mixtures also confirm this finding. Indeed, three to thirteen times increase of conductivity is observed upon the dilution of pure RTIL with acetonitrile. Quite interestingly, the impact of ACN on the conductivity of [BMIM][BF$_4$]/ACN is noticeably larger than on [EMIM][BF$_4$]/ACN, 3-7 times *vs.* 5-13 times. In order to understand this, we should accept that at $x$ (RTIL) < 25% the ionic motion is driven by acetonitrile (Figures 5-6). For instance, at $x$ (RTIL) < 10%, the average diffusion coefficients of [EMIM][BF$_4$]/ACN are 1.1-1.2 times larger than those of [BMIM][BF$_4$]/ACN. In turn, in pure RTILs, this ratio equals to 2.5. It is deduced that high conductivity of the considered systems is achieved due to a high mobility of the molecules of aprotic solvent, rather than to the transport properties of RTILs *per se*. Practically, this trend is extremely important, since allows for using arbitrary RTIL of the imidazolium family for applications. In pure RTILs, conductivity greatly depends on the cation size (Table S1), being noticeably smaller for longer aliphatic tails. Importantly, our simulations predict the maximum of σ at the same narrow composition interval as recent experimental studies.[26] However, quantitative coincidence is not excellent, conductometry data being somewhat smaller than ours. It is quite strange, since the simulated conductivities might be expected smaller due to slower diffusion of the force field model used for ACN, see above. The mentioned experiment also contradicts the most recent data by Lopes *et al.*,[47] whose conductivities, measured for very similar RTIL, 1-butyl-3-methylimidazolium bis(trifluoromethanesulfonyl)-imide, are predictably higher.

Another strong correlation of conductivity is found to be with temperature. Whereas at 323 K the conductivity of [EMIM][BF$_4$]/ACN increases by 3 times upon dilution, at 283 K its increase is 7 times. In the case of [BMIM][BF$_4$]/ACN, a similar trend is found, 5 and 13 times, respectively (Table S1). Therefore, using mixtures instead of pure RTILs is expecially important for electrolyte applications at 283 K and lower temperatures. The correlation of the ionic conductivity with long-range structure (ionic clusters) of RTILs is discussed below (Figures 12-13).

The observed trends are generalized by considering three more RTILs with longer alkyl tails, containing 6, 8, and 10 carbon atoms. The specific density and all transport properties are depicted as a function of the RTIL tail length for 10% and 100% RTIL contents in Figure 8. The elongation of the lyophobic part of the cation leads to the system dynamics retardation, although the interaction energy per atom obviously decreases. This is an important, yet not intuitive, finding, which underlines that each increment of the alkyl tail plays a significant role for the imidazolium-based moieties. In pure RTILs with tails of 8 and 10 carbon atoms, the diffusion coefficients, and consequently, conductivities are negligible (< 0.1 S/m), while the viscosities grow to extra-large values (> 400 cP at 283 K and > 150 cP at 298 K). Remarkably, ACN boosts the ionic transport drastically (Figure 8e), as its content approaches 10 molar percents. For instance, the conductivity of pure [DMIM][BF$_4$] at 283 K is 0.029 S/m, *i.e.* 62 times increase, while in the 10% mixture it is 1.8 S/m, that is only 2.4 times smaller than of [EMIM][BF$_4$] (4.3 S/m). At the elevated temperature (323 K), the trend is not so sharp (26 times increase), but nevertheless very impressing. It largely depends on the fact that 283 K is below the glass transition temperature for the longest-tail RTILs, and the solvation changes their phase to liquid. Meanwhile, the difference between [EMIM][BF$_4$] (the highest conductivity) and the RTILs with larger cations is 2.4 times at most, whereas in the case of pure RTILs it is 19-22 times ([EMIM][BF$_4$] *vs.* [DMIM][BF$_4$]). The position of the conductivity maximum (Figure 7, Tables S1-S2) is invariable for all five RTILs.

Note, 10 molar percents is quite a large concentration corresponding to *ca.* 46 w/w % in the case of [DMIM][BF$_4$]/ACN mixture. This content has an amazingly tiny impact on the density of the systems, which differs by just 10-12 kg/m$^3$, as the number of alkyl carbon atoms grows from 2 to 10. Compare, in the case of pure RTILs the corresponding density decrease is *ca.* 80 kg/m$^3$. It is important to understand that usually the overall density of the system plays a crucial role in the ionic/molecular transport. Similarly, here the density and ionic dynamics are very tightly correlated (Figure 8).

Figure 9 summarizes the influence of ACN at 298 K. Interestingly, the conductivity increase, observed for [EMIM][BF$_4$] and [BMIM][BF$_4$] and discussed above, is much smaller than that for [HMIM][BF$_4$], [OMIM][BF$_4$], and [DMIM][BF$_4$]. For 25% RTIL mixtures, the corresponding difference is somewhat smaller, while for 5 and 10 % it is nearly the same. Analyzing Figures 7-10 jointly, one can conclude that ionic transport of all the investigated imidazolium-based RTILs is driven by the affinity to cosolvent (acetonitrile) rather than by the properties (shape, mass, polarity) of the ionic species themselves. As will be demonstrated below (Figures 12-13), the binary mixtures of imidazolium-based RTILs and ACN form true solutions, thanks to exceptionally favorable interactions of the components. Noteworthy, such behavior is not reported for any other co-solvent including water[47].

Compare, at 298 K, the maximum conductivity of [BMIM][BF$_4$] is 1.9 S/m in butanon, 1.6 S/m in ethanol, 1.1 S/m in dichloromethane, 3.1 S/m in acetone (ACT).[26] Additionally, it may be informative to compare the simulated conductivities of [BMIM][BF$_4$]/ACN with those of [BMIM][TFSI],[47] which are 0.68 S/m in 1-butanol, 2.6 S/m in methanol, 1.5 S/m in dichloroethane, and 4.4 S/m in ACN. Clearly, ACN provides the largest conductivity increase, which can be compared only with that in ACT. In this context, investigation of longer-tailed RTIL/ACT mixtures can be of great interest. ACT and ACN are obviously similar polar solvents, μ(ACN)=3.9 D, μ(ACN)=2.9 D, both possessing insignificantly polar methyl groups.

So far, the [OMIM][BF$_4$]/ACN and [DMIM][BF$_4$] mixtures were not considered for electrochemistry, since it was believed that transport of these massive, bulky cations was unreasonably hindered. However, as we prove in the present work, based on the new, feasible force field for transport properties, in ACN their motion is only moderately slower than that of the short-tailed RTILs. Our findings, therefore, open a new exciting opportunity in tuning electrolyte solutions, consisting of imidazolium-based RTILs. The trends for cations may be expected independent of the anion, if the latter is such a common moiety as PF$_6$ or bis(trifluoromethanesulfonyl)-imide, since they demonstrate generally the same nature of interactions. Overall, the enhancement of σ upon the addition of ACN greatly favors applications of the imidazolium-based RTILs/ACN mixtures in the electrolyte containing devices, including batteries, supercapacitors, solar cells, *etc.*

The $H_{vap}$s of pure components (Figure 10) are very different (about four times), justifying *ca.* 100-fold difference in the viscosity and diffusion and almost two-fold difference in specific density. The $H_{vap}$s of the [BMIM][BF$_4$]/ACN mixtures are slightly higher than those of the [EMIM][BF$_4$]/ACN mixtures, originating from the longer, although lyophobic, alkyl tail. The derivatives of $H_{vap}$ with respect to molar fraction are 1.06 and 0.95 kJ/mol for [BMIM][BF$_4$]/ACN and [EMIM][BF$_4$]/ACN, respectively, whereas the correlation coefficients (>0.999) indicate a tight accuracy of the derived data. One can speculate that the observed behavior of $H_{vap}$ is common for all imidazolium-based ionic liquids, since electrically neutral alkyl tail brings only insignificant contribution to the total interaction energy between RTIL and molecular solvent.

It should be noted that since the experimentally observed volatilities of RTILs and ACN are very different, as well as their boiling points, the compositions of the liquid and vapor phases vary appreciably. Actually, at 298 K the saturated vapor comprises exclusively ACN molecules, whereas ions remain in the liquid phase. Hence, a formal definition given in eq. (1) does not describe an observed phenomenon. In order to compare our results with a physical experiment, they should be corrected considering $U([EMIM][BF_4], \text{ion pair}) = 110$ kJ/mol and $U([BMIM][BF_4], \text{ion pair}) = 103$ kJ/mol. In the meantime, the values in Figure 10 are an important integral measure of the energy, which is experienced by ions in the condensed phase. Naturally, they are directly proportional to viscosity and inversely proportional to conductivity of the $[EMIM][BF_4]/ACN$ and $[BMIM][BF_4]/ACN$ mixtures.

In order to corroborate our assumption about correlation between ionic and molecular diffusion, and consequently, ionic conductivity, and supra-ionic structures, a cluster analysis is performed for varying composition of RTILs (Figures 12-13). It is based on the criteria derived from radial distribution functions (Figure 11) between the strongly interacting sites of the cation and the anion, $g_{CR-B}(r)$. Noteworthy, $g_{CR-B}(r)$ according to the position (0.38 nm), height (ca. 3-5) and shape of the first peak predicts the formation of the ion pairs at $5\% < x$ (RTIL) $< 50\%$. Herewith, the probability of the ion pair formation and the duration of its existence are in direct proportion to the content of ACN. The fact that the amount of ion pairs increases as the general quantity of ions decreases is a remarkable observation, suggesting that ion-molecular interactions stibilize ionic aggregates. In turn, the formation of neutral ionic aggregates implies that conductivity decreases. The inter-ionic RDF also contains a well pronounced second maximum at 0.59 nm with a height of 1.9-2.6. This maximum corresponds to larger ionic aggregates. Surprizingly, no acetonitrile-separated ionic pairs are detected on $g_{CR-B}(r)$, although it was previously shown that ACN creates such species with conventional electrolytes[48]. The shapes of $g_{CR-B}(r)$ for [EMIM][BF$_4$]/ACN and [BMIM][BF$_4$]/ACN are nearly identical. An immediate snapshot in Figure 1 shows a few aggregates which can be formally designated as solvent-separated ion pair. However, $g_{CR-B}(r)$ does not support the stability of such formations. In turn, the absence of the third peak in RDF underlines that ACN molecules screen ion-ionic interactions in [EMIM][BF$_4$]/ACN and [BMIM][BF$_4$]/ACN very efficiently.

As well as RDF, the cluster analysis (Figures 12-13) suggests that the ionic structure regularities are very similar among imidazolium-based RTILs. Firstly, the size of the biggest ionic cluster is in direct proportion to molar fraction. Second, the preferencial aggregate is ion pair, whose formation is more probable in ACN-rich mixtures rather than in RTIL-rich ones. Third, significant percentages of free ions exist only at $x$ (RTIL) = 5 % and $x$ (RTIL) = 10 %. Interestingly, at $x$ (RTIL) = 25 % and $x$ (RTIL) = 50 %, the amount of free ions is comparable with the amount of ion pairs. It indicates that the ion pairs in these mixtures lack stability. As a result, a dynamical equilibrium between ionic aggregates of varying composition is established. It also suggests that mixing with ACN is thermodynamically very favorable for the imidazolium-based RTILs.

The average size of the ionic cluster at 298 K is 1.65, 2.16, 5.13, and 41.86 for 5, 10, 25, and 50 % of [EMIM][BF$_4$] in ACN, respectively. The similar distribution is also observed in the case of [BMIM][BF$_4$] in ACN at 298 K, providing 1.67, 2.16, 4.86, and 19.13 for 5, 10, 25, and 50 %, respectively. The average cluster sizes at different temperatures for both RTILs are summarized in Table 2. Whereas for 5 and 10 % of [EMIM][BF$_4$] and [BMIM][BF$_4$], the cluster size looks temperature independent for both ionic liquids, for 25 and 50% it strongly decreases as temperature increases. This is presumably connected with an enhanced mobility of acetonitrile upon heating towards its boiling point, whose diffusion pace exponentially depends on temperature ($D_{AN}$ = 2.5, 3.3, and 4.3 ($\times 10^{-9}$ m$^2$/s) at 283, 298, and 323 K, respectively).

Another important point is a systematic decrease of the average cluster size in [BMIM][BF$_4$]/ACN as compared to [EMIM][BF$_4$]/ACN, if $x$ (RTIL) > 10 %. For instance, at $x$ (RTIL) = 50% the difference between these two RTILs exceeds two times over the entire temperature range. Provided that $g_{CR-F}(r)$ for both EMIM$^+$ and BMIM$^+$ (Figure 11) are very similar, such a behavior may contradict an intuition. By analyzing the probability of formation *vs.* cluster size (Table 3), we found that BMIM$^+$ and BF$_4^-$ tend to create a slightly larger amount of smaller clusters, while EMIM$^+$ and BF$_4^-$ create larger clusters. At $x$ (RTIL) = 50%, the largest cluster sizes are 240 and 170 ions for [EMIM][BF$_4$] and [BMIM][BF$_4$], respectively. It should be noted that the probability of their formation is negligible (*ca.* 0.01 %). The largest clusters whose probability of existance is larger than 1% are 22 ions for [EMIM][BF$_4$] and 26 ions for [BMIM][BF$_4$]. Compare, the total amount of ions in this simulated mixture (50% RTIL) is 300.

Based on the numerous *ab initio* calculations, it is recognized that the imidazole rings of EMIM$^+$ and BMIM$^+$ possess the same chemical and physical properties, so they differ only by a length of the lyophobic alkyl tail. Remarkably, this tiny distinction leads to a noticeable difference in the cluster sizes but nohow evince itself in the local structure (RDFs). Since the alkyl tails do not exhibit any specific interactions with an environment, the smaller clusters in case of BMIM$^+$ and BF$_4^-$ should be stipulated by a steric factor exclusively. It is confirmed also by an observation that the probability of the lonely ions existence is higher in [BMIM][BF$_4$]/ACN (7.98 %) than in [EMIM][BF$_4$]/ACN (7.46 %). It may be instructive to investigate cluster formation for the imidazolium-based RTILs with longer tails (1-methyl-3-hexylimidazolium, 1-methyl-3-octylimidazolium, 1-methyl-3-decylimidazolium, *etc*) to understand the extent to which chemically inert groups (C$_x$H$_y$) are able to modify ionic structure of this sort of compounds.

An ability of charged particles to create large and long-lived ionic aggregates is one of the principal factors, which determine $\sigma$. The ionic species, belonging to stable aggregate (cluster), tend to move as a whole, that results in a hindered diffusion, and therefore, low conductivity. In the case of the imidazolium-based RTIL/ACN mixtures, the average cluster size is directly proportional to the content of acetonitrile. However, as the content of ACN grows, the total number of the charge carriers decreases. So, the position of the maximum on the conductivity $vs.$ molar fraction plot is determined by the above two factors. Remarkably, the position of the maximum of conductivity (10-25% of RTIL, Figure 7) coincides with a molar fraction of RTIL, where the percentage of single ions exceeds the percentage of ion pairs. In turn, at larger $x$ (RTIL), the ratio between the probability of formation of these clusters is inverse. As the content of RTIL decreases down to 5%, the number of free ions increases, but conductivity nevertheless decreases. Therefore, here $\sigma$ is determined by the total concentration of charge carriers rather than the ionic structure of the respective RTIL. Based on the reported observations, one can speculate that the conductivity maximum is roughly found, where the ratio between free ions and ion pairs is being inversed. Such predictions can be done either using molecular simulation or experimental techniques.

**Conclusions**

To recapitulate, extensive molecular dynamics simulations of the [EMIM][BF$_4$]/ACN, [BMIM][BF$_4$]/ACN, [HMIM][BF$_4$]/ACN, [OMIM][BF$_4$]/ACN, and [DMIM][BF$_4$]/ACN mixtures at 283, 298, and 323 K are reported. Specific densities, radial distribution functions, ionic cluster distributions, heats of vaporization, diffusion constants, shear viscosities, and ionic conductivities and their correlations are discussed. It is found that the dilution of the considered imidazolium-based RTILs with acetonitrile allows for conductivity increase by more than 50 times for long-tailed RTILs and more than 10 times for short-tailed RTILs.

The ionic conductivity is a complex function of the mixture content and external conditions, whose is not trivial without atomistic-precision description of the substance/mixture. Nevertheless, its maximum correlates well with a long-range ionic structure (ionic clusters). Namely, the maximum is observed at the molar fraction of RTIL, where the number of free ions exceeds the number of ion pairs. Remarkably, the dilution-driven conductivity increase is significantly larger for [DMIM][BF$_4$], than for [EMIM][BF$_4$], since the first one forms smaller ionic clusters. The corresponding increase for [DMIM][BF$_4$], [OMIM][BF$_4$], and [HMIM][BF$_4$] is even higher, probably because of the larger difference between $D$(ACN) and $D$(pure RTILs). Since the system dynamics in 5 and 10% RTIL systems are driven by the molecular cosolvent, the motion of all ions tends to approach that of ACN. Accordingly, the addition of ACN exponentially increases diffusion and decreases shear viscosity of the mixtures. The radial distribution functions show that molecular solvent stabilizes ion pairs, whereas the ionic clusters of larger size are ruined, as suggested by our cluster analysis. The reliability of the reported data is justified by the available experimental densities and viscosities of these mixtures at 298 K.

Despite all virtues of the considered systems, using ACN as a component of electrolyte solution arises an important safety question. Indeed, ACN exhibits a moderate toxicity and flammability. This issue deserves a comprehensive investigation[49] using both simulation and experimental approaches. Meanwhile, based on a very good miscibility of ACN with the imidazolium-based RTILs one can assume that physical properties causing its flammability and toxicity, *e.g.* volatility, are significantly modified. Considering the dependence of self-diffusion on molar fraction, an ability of ACN to evaporate from the RTIL/ACN mixtures is to be thoroughly reconsidered.

## Acknowledgments

Vitaly Chaban is grateful to Julianne Green (University of Rochester) for valuable considerations. The research is supported in part by the NSF grant CHE-1050405. Iuliia

Voroshylova acknowledges the FCT for the research grant (project PTDC/EQU-FTT/104195/2008).

**Supporting Information**

The supporting information includes Tables S1-S2, listing the viscosities and conductivities of the imidazolium-based RTILs/ACN binary mixtures over a wide range of compositions.

**TABLES**

Table 1. The number of interacting sites per system consisting of the five imidazolium-based RTILs and acetonitrile. Each of these systems is consequently equilibrated and simulated at 283, 298, and 323 K

| $x_1$ ($n_1/n_2$)* | [EMIM][BF$_4$] | [BMIM][BF$_4$] | [HMIM][BF$_4$] | [OMIM][BF$_4$] | [DMIM][BF$_4$] |
|---|---|---|---|---|---|
| 5 (15/285) | 2070 | 2160 | 2250 | 2340 | 2430 |
| 10 (30/270) | 2340 | 2520 | 2700 | 2880 | 3060 |
| 25 (75/225) | 3150 | 3600 | 4050 | 4500 | 4950 |
| 50 (150/150) | 4500 | 6300 | — | — | — |
| 75 (225/75) | 5850 | 7200 | — | — | — |
| 90 (270/30) | 6660 | 8280 | — | — | — |
| 100 (300/0) | 7200 | 9000 | 10800 | 12600 | 14400 |

* $x_1$ denotes a molar percentage of RTIL in a mixture, while $n_1$ and $n_2$ are the numbers of ion pairs and ACN molecules, respectively.

Table 2. The average cluster sizes of $[EMIM^+]_n[BF_4^-]_m$ and $[BMIM^+]_n[BF_4^-]_m$ in their mixtures with ACN derived from MD simulations at 283, 298, and 323 K. Note, the shown cluster size is a sum of counterions, *i.e. n+m*.

| $x$ (RTIL),% | $[EMIM][BF_4]$ | | | $[BMIM][BF_4]$ | | |
|---|---|---|---|---|---|---|
| | 283 K | 298 K | 323 K | 283 K | 298 K | 323 K |
| 5 | 1.59 | 1.65 | 1.61 | 1.63 | 1.67 | 1.71 |
| 10 | 2.23 | 2.16 | 2.22 | 2.22 | 2.16 | 2.17 |
| 25 | 5.69 | 5.13 | 5.14 | 4.95 | 4.86 | 4.42 |
| 50 | 45.30 | 41.86 | 34.56 | 21.24 | 19.13 | 16.49 |

Table 3. The probabilities of the formation of [EMIM$^+$]$_n$[BF$_4^-$]$_m$ and [BMIM$^+$]$_n$[BF$_4^-$]$_m$ ionic clusters in their equimolar mixtures with ACN derived from MD simulations at 323 K

| Cluster size, (number of ions) | Probability, % | | Cluster size, (number of ions) | Probability, % | |
|---|---|---|---|---|---|
| | [EMIM][BF$_4$] | [BMIM][BF$_4$] | | [EMIM][BF$_4$] | [BMIM][BF$_4$] |
| 1-10 | 41.70 | 54.88 | 91-100 | 2.39 | 0.50 |
| 11-20 | 14.14 | 19.10 | 101-110 | 2.00 | 0.29 |
| 21-30 | 8.44 | 9.78 | 111-120 | 1.76 | 0.16 |
| 31-40 | 5.97 | 5.81 | 121-130 | 1.50 | 0.13 |
| 41-50 | 4.66 | 3.49 | 131-140 | 1.14 | 0.12 |
| 51-60 | 3.97 | 2.42 | 141-150 | 1.09 | 0.07 |
| 61-70 | 3.45 | 1.57 | 151-160 | 0.79 | 0.00 |
| 71-80 | 2.71 | 0.98 | 161-170 | 0.74 | 0.04 |
| 81-90 | 2.60 | 0.66 | 171-240 | 0.95 | 0.00 |

**FIGURE CAPTIONS**

Figure 1. (Color online) The simulated particles, 1-ethyl-3-methylimidazolium$^+$ (EMIM$^+$), 1-butyl-3-methylimidazolium$^+$ (BMIM$^+$), tetrafluoroborate$^-$ (BF$_4^-$), acetonitrile (ACN), and the simulated system of the equimolar mixture of [BMIM][BF$_4$] and ACN.

Figure 2. (Color online) (a) The specific density and (b) the excess molar volumes of [EMIM][BF$_4$]/ACN and [BMIM][BF$_4$]/ACN mixtures computed at 283 K (red circles), 298 K (green squares), and 323 K (blue triangles) as a function of molar fraction. The connecting lines are present to guide an eye.

Figure 3. (Color online) The parameters $\rho_0$ and $a$ in $\rho = \rho_0 + a \cdot T$, describing temperature dependence of the specific density of [EMIM][BF$_4$]/ACN (red circles) and [BMIM][BF$_4$]/ACN (green squares) mixtures as a function of molar fraction. The connecting lines are present to guide an eye.

Figure 4. (Color online) The shear viscosity of the [EMIM][BF$_4$]/ACN and [BMIM][BF$_4$]/ACN mixtures computed at 283 K (red circles), 298 K (green squares), and 323 K (blue triangles) as a function of molar fraction. The uncertainty of any simulated value does not exceed 20%. The connecting lines are present to guide an eye.

Figure 5. (Color online) The diffusion constants of cation (EMIM$^+$), anion (BF$_4^-$), solvent molecules (ACN) and average diffusion constants of the [EMIM][BF$_4$]/ACN mixtures computed at 283 K (red circles), 298 K (green squares), and 323 K (blue triangles) as a function of molar fraction. The uncertainty of any simulated value does not exceed 10%. The connecting lines are present to guide an eye.

Figure 6. (Color online) The diffusion constants of cation (BMIM$^+$), anion (BF$_4^-$), solvent molecules (ACN) and average diffusion constants of the [BMIM][BF$_4$]/ACN mixtures computed at 283 K (red circles), 298 K (green squares), and 323 K (blue triangles) as a function of molar fraction. The uncertainty of any simulated value does not exceed 10%. The connecting lines are present to guide an eye.

Figure 7. (Color online) The ionic conductivity of the [EMIM][BF$_4$]/ACN and [BMIM][BF$_4$]/ACN mixtures computed at 283 K (red circles), 298 K (green squares), and 323 K (blue triangles) as a function of molar fraction. The uncertainty of any simulated value does not exceed 15%. The connecting lines are present to guide an eye.

Figure 8 (Color online) The ionic conductivity (a, b), shear viscosity (c, d), diffusion coefficient (e, f), and specific density (g, k) as a function of the number of carbon atoms forming the alkyl tail in 10% RTIL/ACN mixtures (a, c, e, g) and pure RTILs (b, d, f, k) at 283 K (red circles), 298 K (green squares), and 323 K (blue triangles). The connecting lines are present to guide an eye.

Figures 9. The ionic conductivity increase upon adding acetonitrile at 298 K, expressed as a ratio of conductivity of 5, 10, and 25% RTIL/ACN mixtures, $\sigma$ (RTIL/ACN), to conductivity of pure RTIL, $\sigma$ (pure RTIL), as a function of the number of carbon atoms forming the alkyl tail.

Figure 10. (Color online) The heat of vaporization of the [EMIM][BF$_4$]/ACN (red circles) and [BMIM][BF$_4$]/ACN (green squares) mixtures computed at 298 K as a function of molar fraction. The connecting lines are present to guide an eye.

Figure 11. (Color online) The radial distribution function, $g_{CR-B}(r)$, between the cation and the anion in the [EMIM][BF$_4$]/ACN (top) and [BMIM][BF$_4$]/ACN (bottom) mixtures as a function of molar fraction. The legend shows the molar fractions of RTIL in the mixtures.

Figure 12. The probability of the formation of ionic clusters in the [EMIM][BF$_4$]/ACN mixtures at 283, 298, and 323 K.

Figure 13. The probability of the formation of ionic clusters in the [BMIM][BF$_4$]/ACN mixtures at 283, 298, and 323 K.

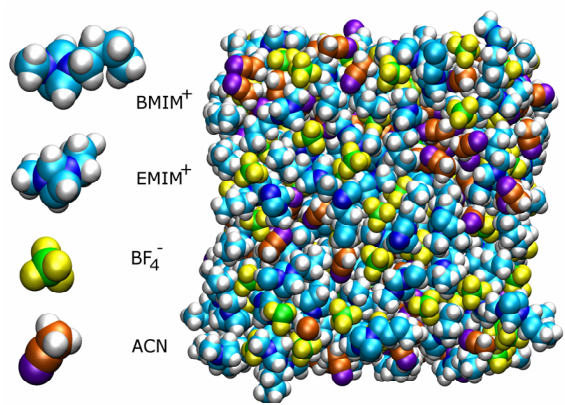

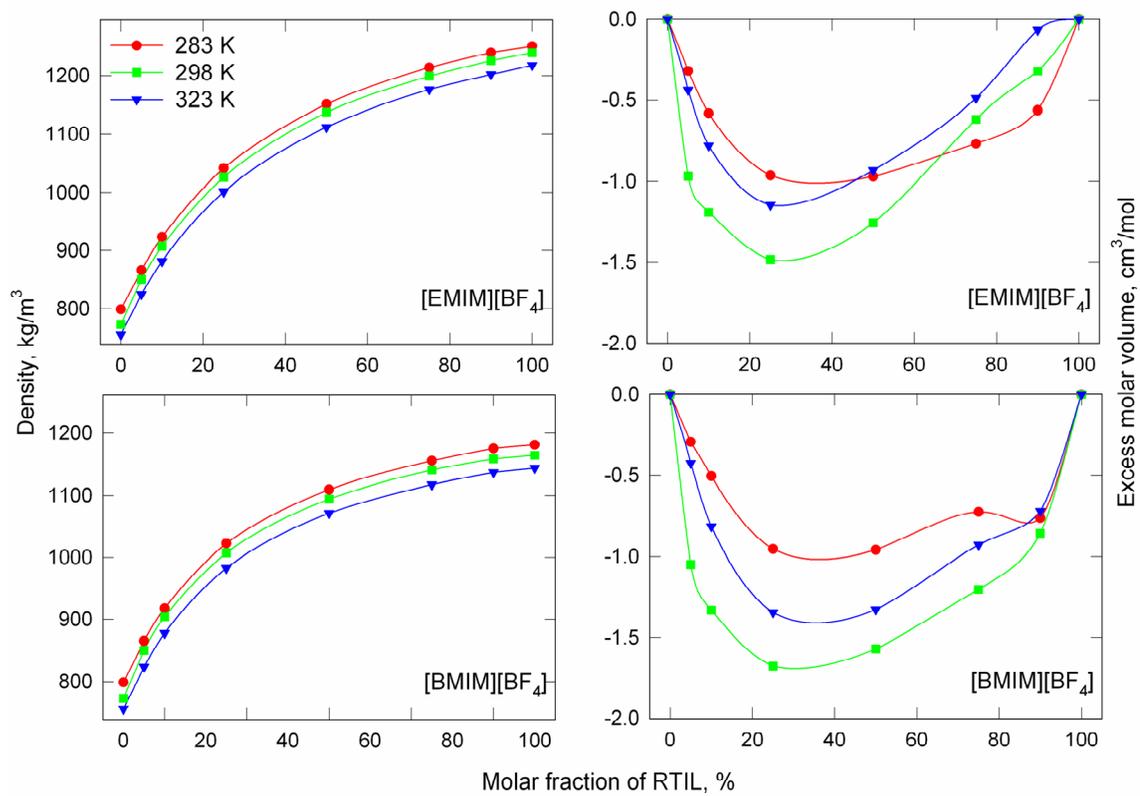

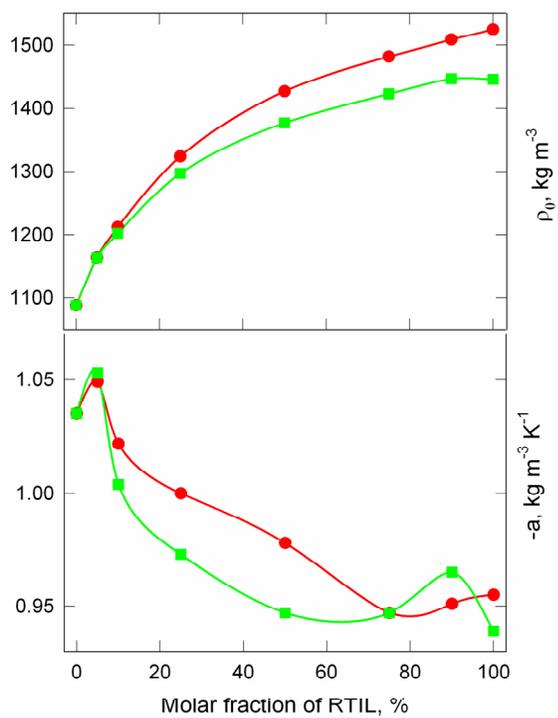

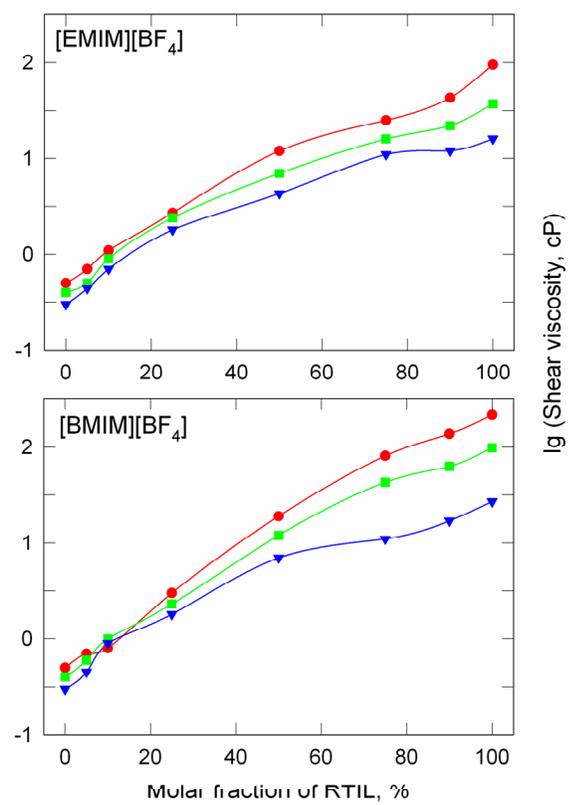

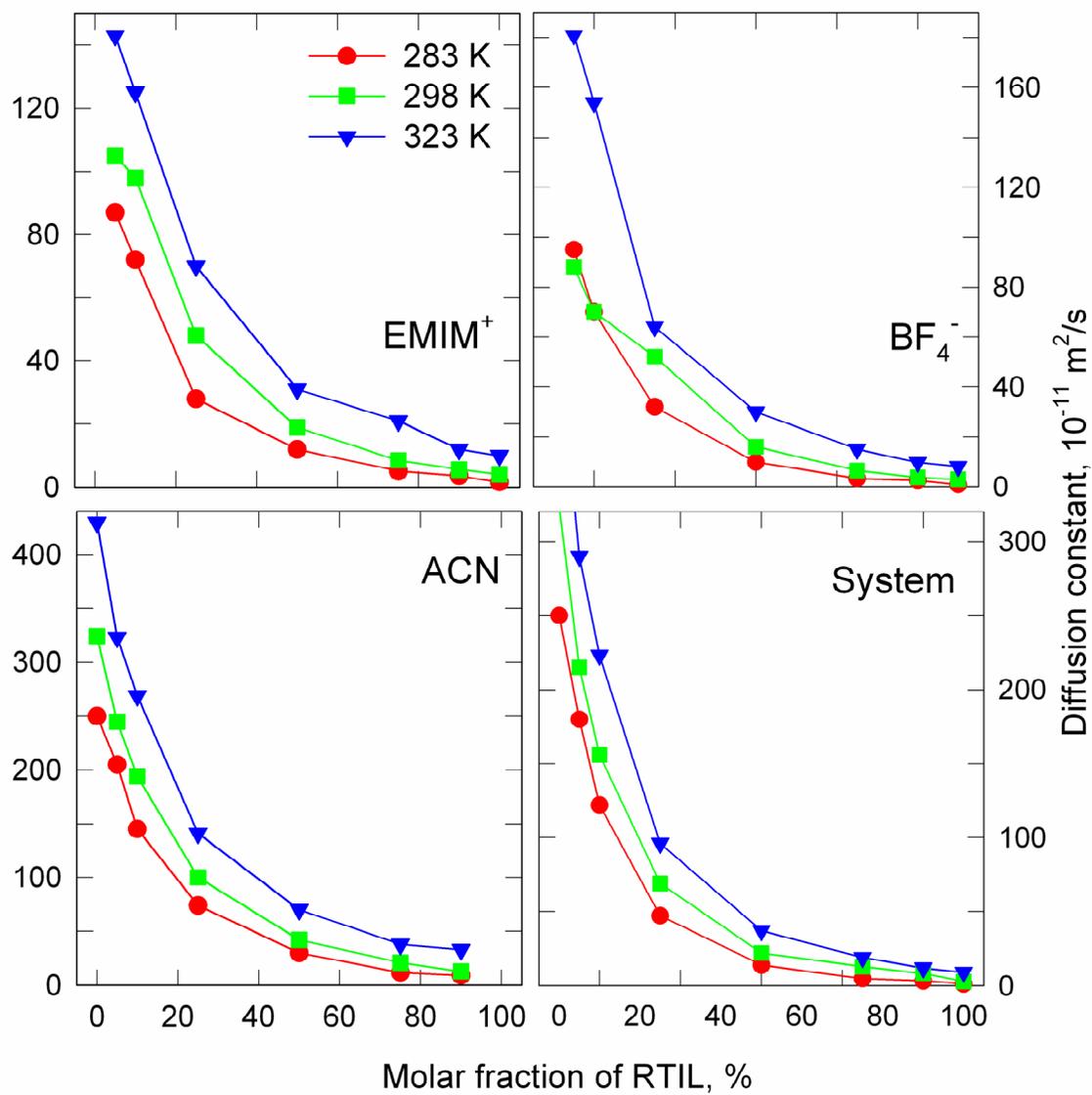

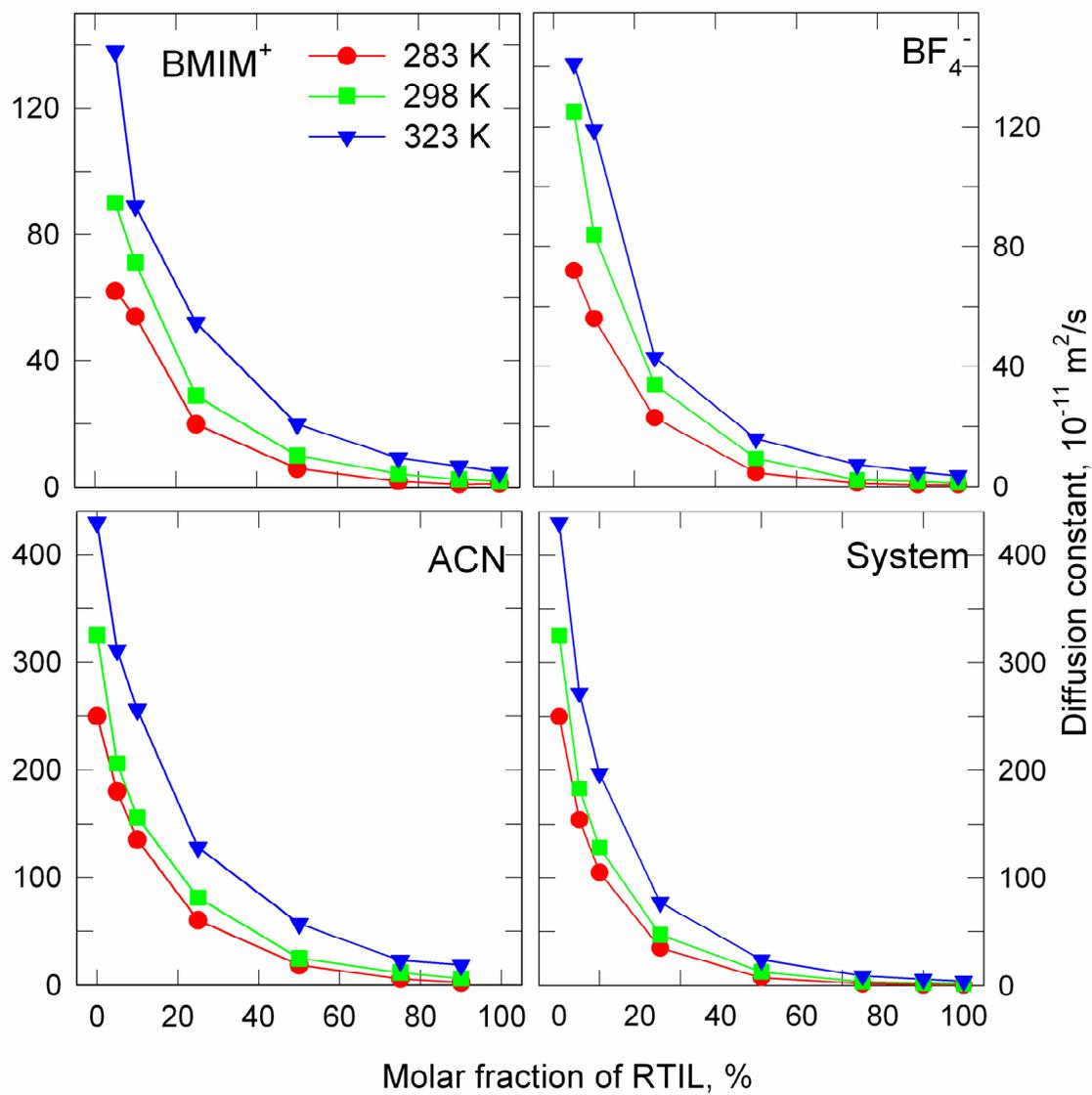

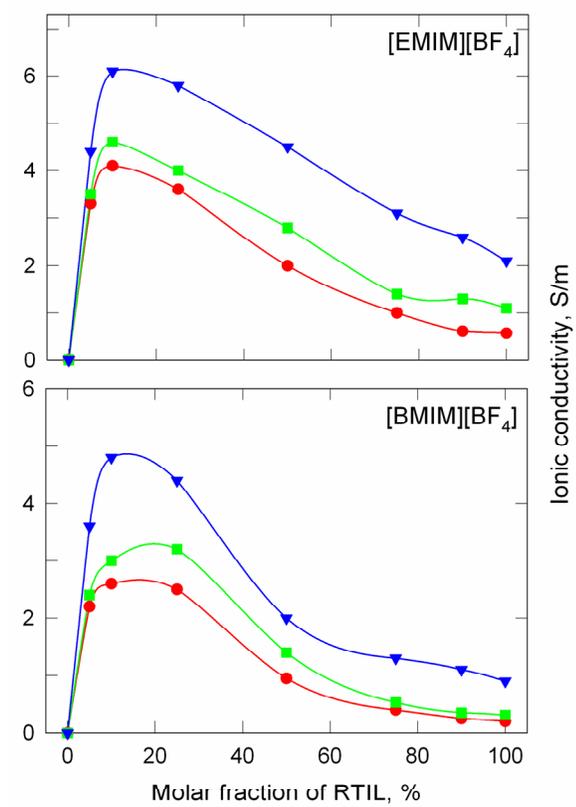

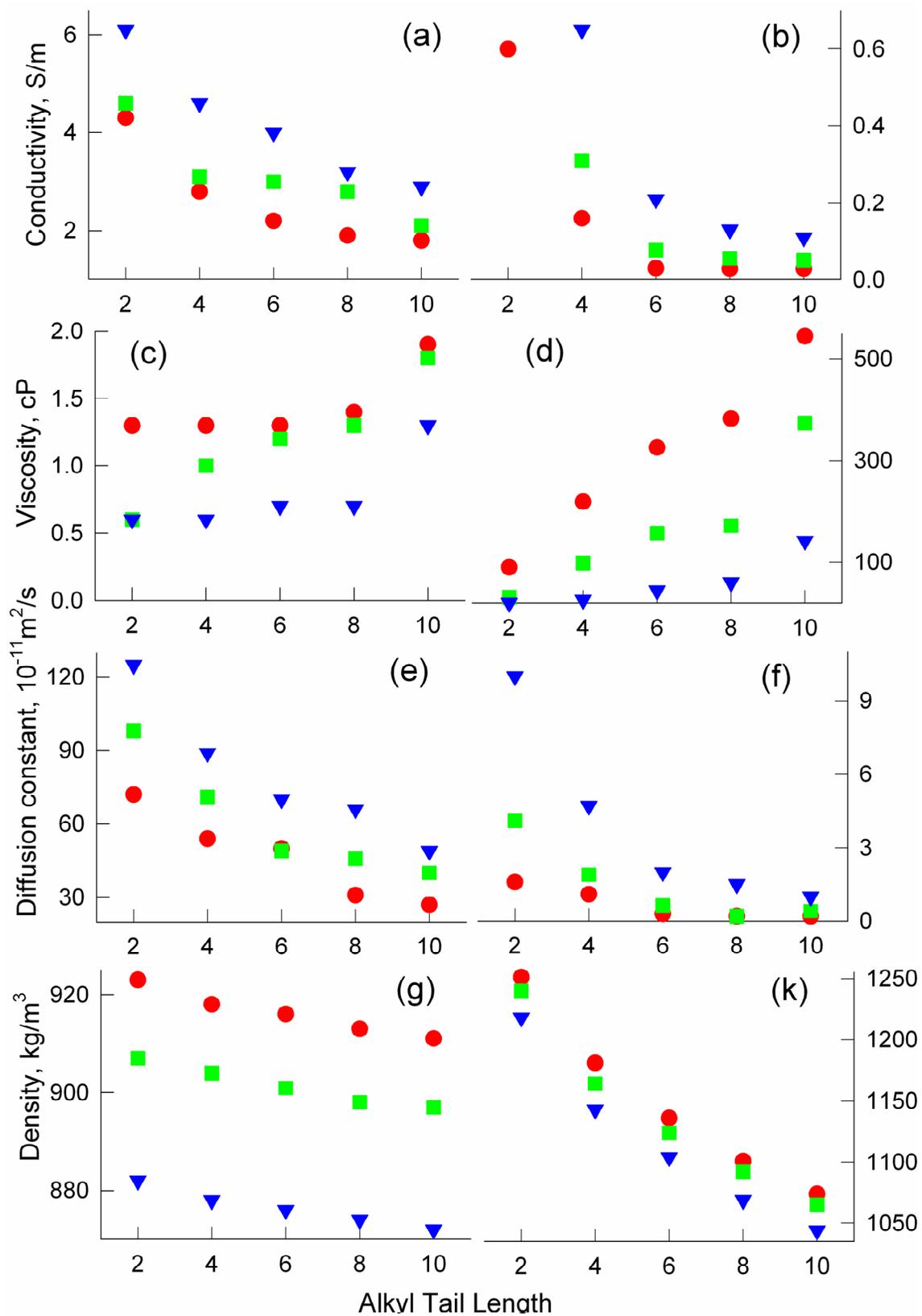

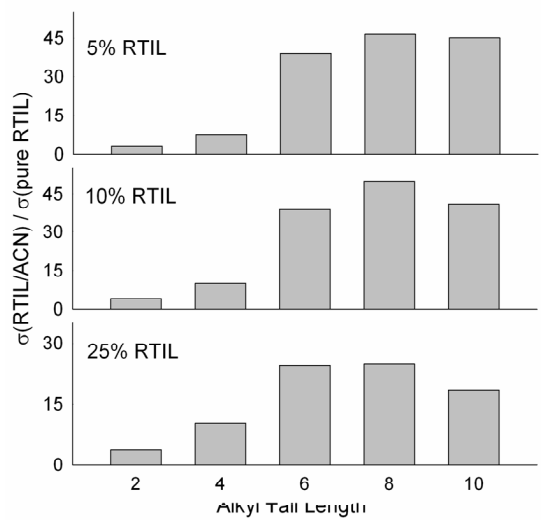

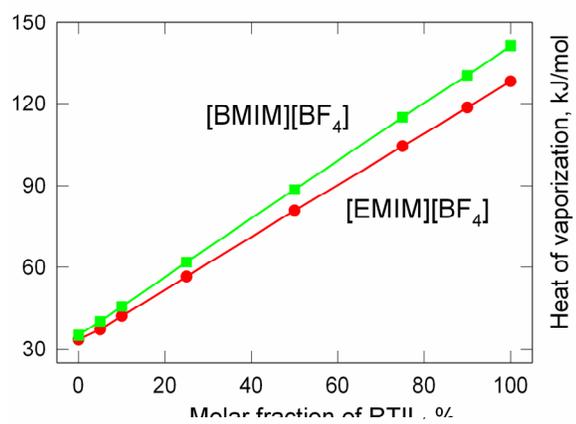

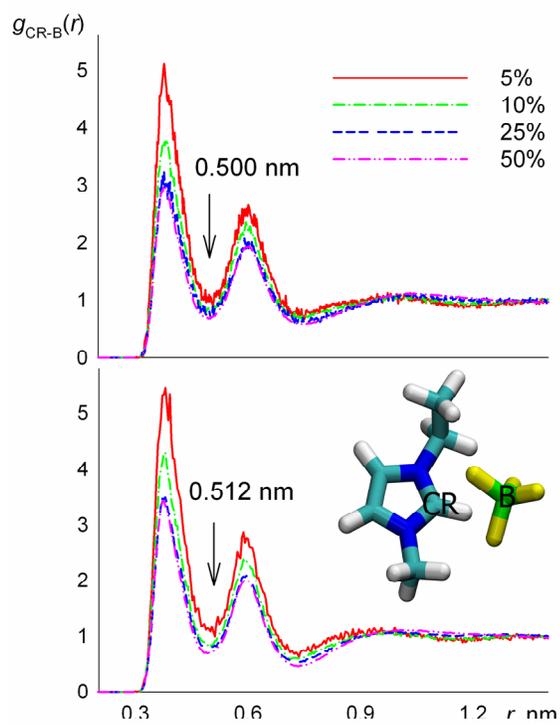

$g_{CR-B}(r)$

0.500 nm

5%
10%
25%
50%

0.512 nm

0.3    0.6    0.9    1.2    $r$, nm

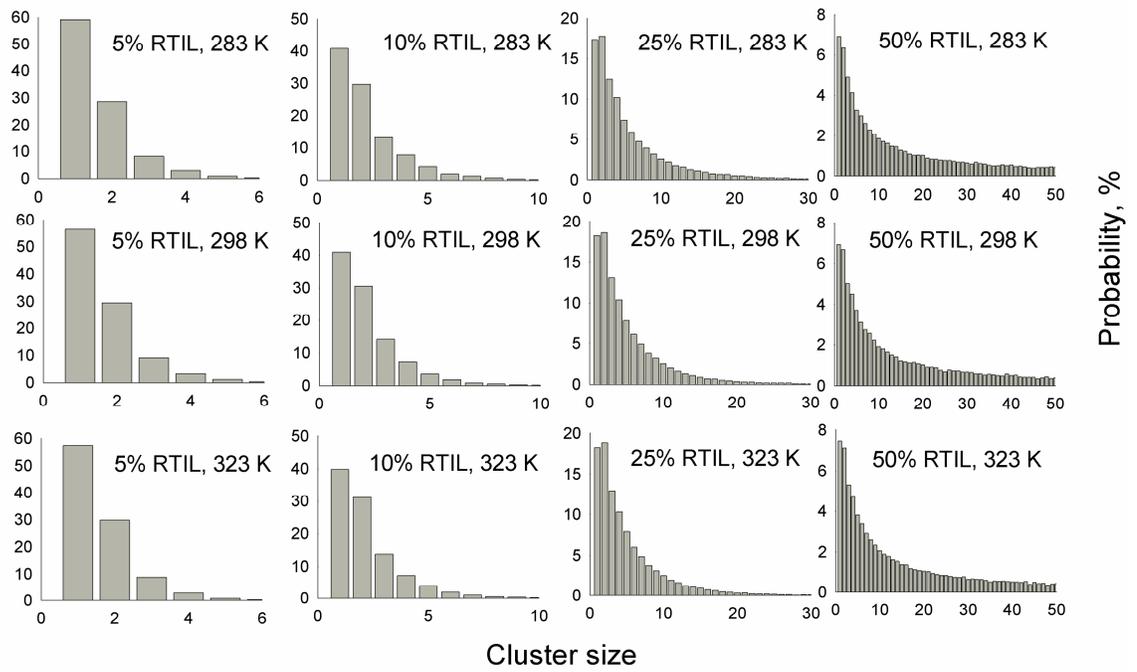

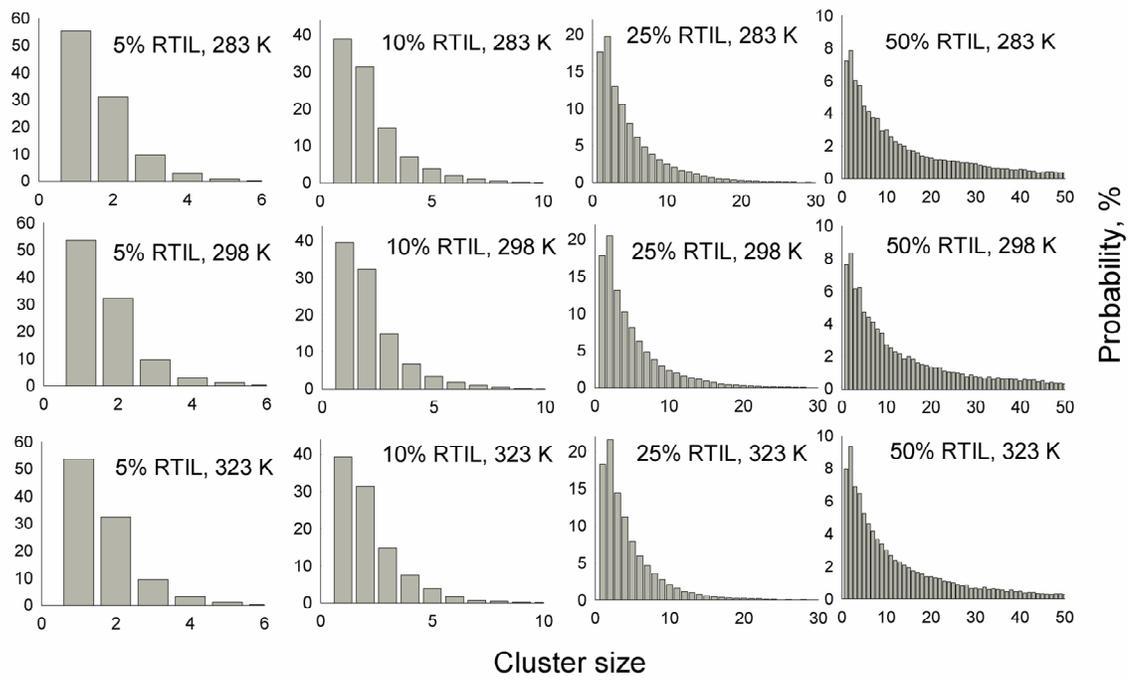

Cluster size

Probability, %